\documentclass[aps,prl,twocolumn, superscriptaddress, showpacs]{revtex4-1}
\usepackage{graphicx}

\begin{document}

\title{Quasi-phase-matching high-harmonic radiation using chirped THz pulses}

\author{Katalin Kov\'acs}
\email[]{kkovacs@itim-cj.ro}
\affiliation{Department of Optics and Quantum Electronics, University of Szeged, HU-6720 Szeged, Hungary}
\affiliation{National Institute for R\&D of Isotopic and Molecular Technologies, RO-400293 Cluj-Napoca,
Romania}
\author{Emeric Balogh}
\affiliation{Department of Optics and Quantum Electronics, University of Szeged, HU-6720 Szeged, Hungary}
\author{J\'anos Hebling}
\affiliation{Department of Experimental Physics, University of P\'ecs, HU-7624 P\'ecs, Hungary}
\author{Valer To\c sa}
\affiliation{National Institute for R\&D of Isotopic and Molecular Technologies, RO-400293 Cluj-Napoca,
Romania}
\author{Katalin Varj\'u}
\affiliation{Department of Optics and Quantum Electronics, University of Szeged, HU-6720 Szeged, Hungary}

\date{\today}

\begin{abstract}
High-order harmonic generation in the presence of a chirped
THz pulse is investigated numerically with a complete 3D non-adiabatic
model. The assisting THz pulse illuminates the HHG gas cell laterally
inducing quasi-phase-matching. We demonstrate that it is possible to compensate the phase
mismatch during propagation and extend the macroscopic cutoff of a propagated
strong IR pulse to the single-dipole cutoff.
We obtain two orders of magnitude increase in the harmonic efficiency
of cutoff harmonics ($\approx$170 eV) using a THz pulse of constant wavelength, and a further factor of 3 enhancement when a chirped THz pulse is used.
\end{abstract}


\maketitle

High-order harmonic generation (HHG) in a noble gas medium is currently
the widely used method to produce coherent radiation in the XUV and
soft X-ray regime.
The elementary laser-atom interaction leading to HHG is understood via the three-step model \cite{3step}, but the complete description of the HHG process requires considering the elementary laser-atom interaction together with the macroscopic aspects of laser and harmonic field propagation in the ionized gaseous medium. The temporal, spectral and spatial distortions of the fundamental pulse during propagation result in a varying intensity and phase which strongly influence the produced harmonic radiation on both the single-atom and the macroscopic level \cite{gaarde}.
%(we refer to \cite{gaarde} for a review on macroscopic aspects of HHG).
% if we need further shortening: delete the next sentence.
The ultimate goal in an experiment is to obtain
intense, coherent XUV or soft X-ray radiation which in turn are useful in the designed applications
like molecular imaging or pump-probe experiments \cite{KrauszIvanov}.

\begin{figure}
\includegraphics[width=7cm]{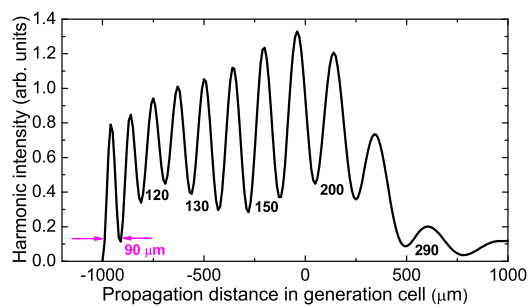}
\caption{\label{fig:h111onax} On-axis variation of the H111 along the propagation direction generated by the IR pulse. The width of several Maker fringes = $2L_{coh}$ is indicated in $\mu$m, other parameters are in the text.}
\end{figure}

In Fig. \ref{fig:h111onax} we show a typical case of phase mismatch (PMM) of the harmonic radiation along the HHG cell manifesting in a quasi-periodic oscillation of the harmonic intensity named Maker fringes \cite{Maker}.
In order to improve phase matching (PM) condition, a number of quasi-phase-matching
(QPM) techniques have been developed to at least partially
compensate for the PMM arising during propagation between
the source and generated harmonics. There are currently several typical
QPM techniques relying  either on periodic modulation of the generating field strength: propagation in diameter-modulated capillaries \cite{capillary}; HHG assisted by a periodic
static electric field (or other periodic waveform e.g. sawtooth)
\cite{biegert,sawtooth};
QPM induced using two counter-propagating pulses \cite{counterprop1,counterprop2};
QPM through multi-mode excitation in a waveguide \cite{multimode}.  An alternative QPM arrangement relies on periodic off-switching of generation in a multi-jet configuration \cite{multijet_seres,multijet_tosa}.

\begin{figure}
\includegraphics[width=4.5cm]{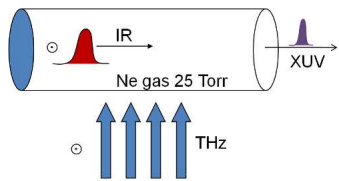}
\caption{\label{fig:sketch}(color online) Schematic representation of the configuration where
the assisting THz pulse propagates in perpendicular direction relative
to the main IR beam which propagates from left to right. Both fields are linearly polarized in the direction perpendicular to the figure plane.}
\end{figure}

We present here a new scheme, inspired by the idea described
in \cite{biegert} where the HHG gas cell was placed in a static
electric field periodically modulated along propagation direction;
a pulsed CO$_2$ laser and an amplitude mask were proposed to create $\approx$ 7MV/cm static electric field
periodically distributed along the HHG cell. In our opinion this configuration was unrealistic because: (1) the CO$_2$ laser pulse with $\approx 10 \mu$m wavelength
cannot be considered as static electric field when compared to the $\approx$100 $\mu$m mask periodicity (2 mm cell length), and (2) producing 7 MV/cm static electric field is not feasible experimentally.
Yet, it is possible to overcome the shortcomings of the QPM configuration presented in \cite{biegert}: we use a THz field imposing a
proper time evolution to create QPM conditions. The advantages
of this new scheme are multiple: strong THz pulses can be generated by
optical rectification of femtosecond laser pulses, and as
such the resulted THz pulse is inherently synchronized in time with
the laser pulse generating the harmonics; fine control in space and time is available
by tuning the wavelength, peak amplitude, chirp rate, initial
phase, thus the shape of the THz pulse can be adapted in order to optimize
QPM and harmonic yield, and there is no need to apply any amplitude mask. We show a schematic representation of a
possible geometric configuration in Fig. \ref{fig:sketch}.
%For experimental realization a standing wave configuration might be more favorable.

The THz field modifies the electron trajectories and hence the single-atom dipole phase (depending on return time and action).
As the generating IR pulse propagates along the gas cell, the temporal
variation of the THz pulse, which is simultaneously present with the IR, translates to a spatial modulation
of the HHG conditions along {\it z}.
According to our results presented below assisting pulses of a few (or few tens) THz frequency and a few MV/cm peak electric field are required for the enhancement of HHG. In this wavelength range the highest electric field already presented is 1.2 MV/cm \cite{hirori} achieved with THz pulses having 1 $\mu$J energy obtained using a tilted pulse front setup \cite{hebling}. Recently THz pulses with 125 $\mu$J energy were generated
and production of THz pulses with more than 10 mJ were predicted \cite{fulop}. These were single-cycle THz pulses, but it is expected that using quasi-sinusoidal intensity modulated pump laser pulses \cite{chen} multicycle THz pulses with similar energy can be generated resulting up to 10 MV/cm field strength in 10 ps long THz pulse on minimum 3 mm diameter spot.

The numerical model in which the THz assisted HHG generation is implemented
is based on a complete 3D non-adiabatic model \cite{tosa2003},
further developed to two-color HHG using
arbitrary wavelength co-propagating pulses \cite{jstqe2011}. The THz
assisted HHG has also been studied in the configuration where the
two pulses propagate in the same direction \cite{imre_pra2011}. For the current work
the model has been extended to describe
pulses propagating perpendicularly.
%\cite{otherpaper}.
% Do we need at all reference to the follow-up paper???
(i) The propagation equation for the IR pulse is solved accounting for diffraction, dispersion (on neutrals and plasma), absorption and nonlinear Kerr effect; (ii) nonlinear dipole response is calculated in the strong field approximation (SFA) \cite{lewenstein}; (iii) the propagation equation for the generated harmonics is solved.
In order to gain a more intuitive insight into the physics
behind the effect of the modulating THz pulse, saddle-point calculations were also carried out providing the phases of the single-atom dipole and the total harmonic field.

We present a case study to demonstrate the new QPM
scheme with THz assisted HHG. We start from a standard HHG system by an IR pulse while looking for the optimal configuration
of the THz pulse which induces the best possible QPM in a selected harmonic range. The particular conditions
for the HHG are the following: the IR pulse has 800 nm central wavelength,
20 fs pulse duration, 0.2 mJ pulse energy, focused with a 20
cm focal length mirror resulting in $8\times10^{14}$W/cm$^{2}$ peak
intensity and 25 $\mu$m beam waist. The HHG cell is 2 mm long filled with 25 Torr
Ne and begins 1 mm before the laser focus. All distances throughout
this paper are measured from the laser focus. The single-dipole cutoff order
is $q=111$ (based on the relation $I_p+3.17U_p$) corresponding to 172 eV and QPM is optimized for
this spectral range.

Fig. \ref{fig:h111onax} illustrates the on-axis intensity variation
of the 111th harmonic (H111) along the propagation axis as generated with the 800 nm pulse alone. We can see the experimentally observable
high-contrast Maker fringes,
%the clear quasi-periodic modulation
a typical result of PMM in nonlinear phenomena.
%The length of an oscillation period corresponds to twice the coherence length ($2L_{coh}$).
On the figure we also
observe that PM conditions vary along the cell, the main causes being the intensity and phase changes of the generating pulse as it propagates.
In particular, the width of the fringes
%period
increases from 90 to 290 $\mu$m during 2 mm of propagation.
The actual period of the high-harmonic intensity modulation and its variation is determined by the focusing conditions of the IR pulse. We use the THz field to modulate HHG conditions along the cell and compensate for PMM. We note here that obviously, looser focusing provides longer coherence lengths, but the here proposed arrangement can be equivalently applied for longer cells with longer THz wavelengths.

\begin{figure}
\includegraphics[width=7.5cm]{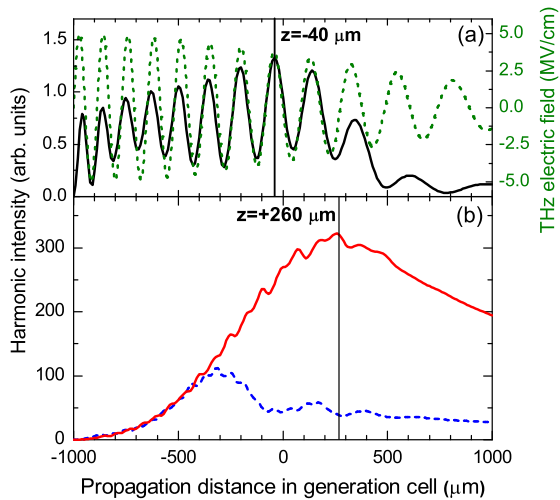}
\caption{\label{fig:chirped}(color online)
(a) Black solid line: THz-free variation of H111 intensity. Green dotted line: $E_{THz}(z)$ visualizes the correspondence with the $2L_{coh}(z)$.
(b) Red solid line: On-axis variation of the H111 intensity along the propagation direction when a chirped THz pulse
is applied. Blue dashed line: Same when the THz pulse has constant $\lambda=120$ $\mu$m.
Vertical lines indicate the positions of maximum harmonic yield.}
\end{figure}

From the results shown in Fig. \ref{fig:h111onax} it is straightforward that for optimal QPM the 
wavelength
%periodicity
of the modulation needs to be varied, a requirement that has been observed in \cite{okeeffe}.
We therefore aim at producing a quasi-periodic spatial modulation of the generating conditions via the temporal modulation of the wavelength (chirp) of our THz pulse.
A QPM effect of the perpendicularly propagating THz field is expected when the modulation caused by the field 
matches the Maker fringes observed
%has the same periodicity as the modulation present
in the IR only case \cite{biegert}. To verify this concept, we studied the effect of a THz field of fixed wavelength ($120\mu$m, corresponding to 2.5 THz) on HHG yield shown in Fig. \ref{fig:chirped}(b) with blue dashed curve. The black solid curve in the top part of the figure illustrates the oscillation of harmonic intensity for the IR only case replotted from Fig. \ref{fig:h111onax} in the same arbitrary units.
%We observe that QPM works in the region where the periodicity of the spatial modulation matches the THz wavelength.
We note here, that the THz amplitude and delay have been chosen appropriately.
%\cite{otherpaper}.

For more efficient QPM we use a chirped THz pulse such that the
wavelength variation of the THz pulse along the cell best matches
the fringe structure
%periodicity
of the chosen harmonic: $\lambda_{THz}(z)\approx2L_{coh}(z)$.
In this case study
we designed a linearly chirped THz pulse that follows nicely the field oscillations, as illustrated in Fig. \ref{fig:chirped}(a) with green dashed curve. The
parameters of the perpendicular THz field are: Gaussian pulse profile with 110 $\mu$m
initial wavelength, $-9\times10^{-7}$fs$^{-2}$ chirp rate resulting 180 $\mu$m average wavelength, 5 MV/cm peak amplitude, 8 ps duration and 4 mm beam waist.
For these parameters the red solid line in Fig. \ref{fig:chirped}(b) indicates very efficient QPM for H111 up to $\approx$1.2 mm in the cell.
We note here that at the end of the cell H111 emission is reduced due to divergence and plasma defocusing of the beam, and the THz field walks off the H111 IR-only intensity modulation pattern leading to loss of QPM.
Comparing the highest intensities along the cell obtained with and without the THz field (realizable experimentally by shortening the cell or moving the focus position) shows a $\approx$300 times yield enhancement.

\begin{figure}
\includegraphics[width=7cm]{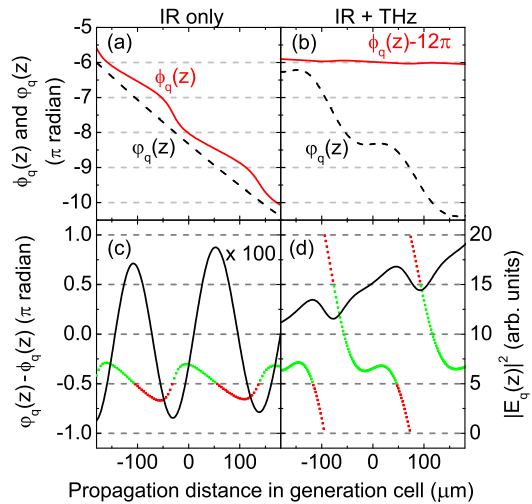}
\caption{\label{fig:phases} (color online) Variation of the H111 dipole phase and total harmonic phase without (a) and with THz field present (b). Red-green dots:  $\varphi_q(z)-\phi_q(z)$ phase difference with scale on the left. Black line: H111 intensity $|E_q(z)|^2$, scale on the right. (c) IR-only case; (d) THz-assisted case.}
%corresponding to the central part of Fig. \ref{fig:chirped}.}
\end{figure}

Optically ionized electrons traveling in the laser field are responsible for HHG, thus both amplitude and phase of the high harmonics are very sensitive to the exact shape of the electric field.
The total harmonic field along {\it z} is the sum of successive dipole emissions:
\begin{equation}
E_q(z)\cdot e^{i\phi_q(z)}=\int_0^z \mathbf{d}^z_q(z') dz'=\int_0^z a_q(z')\cdot e^{i\varphi_q(z')}dz',
\end{equation}
where $E_q(z)$ and $\phi_q(z)$ are amplitude and phase of the q$^{th}$ order HH field accumulated till $z$ and the dipole phase $\varphi_q(z)=q\varphi_{IR}(z)+\Phi_{at}(z)$ is described in a traveling frame moving with $c$.
The applied THz pulse amplitude is weak compared to the generating IR ($<$1\%), thus its main effect is the modulation of the accumulated phase of the electron while traveling in the field ($\Phi_{at}(z)$). Efficient QPM is explained at the harmonic dipole level.
In particular, for H111 the dipole amplitude variation $a_q(z)$ did not show any correlation with $E_q(z)$, instead we found strong correlation with the phase. Using saddle-point calculations for short-to-cutoff trajectories in the central half-cycle of the IR pulse
%generating this order
we studied the $\varphi_q(z)$ and $\phi_q(z)$ variations, and compared the phase difference $\phi_q(z)-\varphi_q(z)$ with the total harmonic intensity $|E_q(z)|^2$. The results are shown in Fig. \ref{fig:phases} for the central part of the cell when HH field is built with IR pulse only (a, c) and when assisted with chirped THz pulse (b, d).
%We note here that the saddle-point calculations were performed for the short-to-cutoff trajectories in the central half-cycle of the IR pulse only, hence the slight difference between Fig. \ref{fig:chirped}a. and Fig. \ref{fig:phases}a.
As expected, while $|\phi_q(z)-\varphi_q(z)|<\pi/2$ the successive dipole emissions add up constructively and result in overall increase of the total harmonic field. The effect of the THz field is that it modulates the dipole phase such to keep $|\phi_q(z)-\varphi_q(z)|$ within $\pi/2$ over longer propagation distances and pass through destructive zones in much shorter regions.

\begin{figure}
\includegraphics[width=7cm]{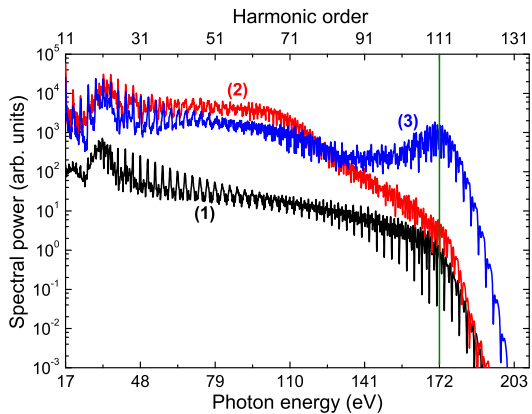}
\caption{\label{fig:spectrum2800} (color online) Logarithmic plot of HH spectra. Black: harmonic spectrum generated with IR
only after 20 $\mu$m propagation. Red: IR only spectrum at the position of max. H111 intensity (-40 $\mu$m).
Blue: spectrum obtained with THz assisted HHG at the position of max. H111 intensity (+260 $\mu$m).
THz pulse was linearly chirped such to best match the on-axis periodicity of H111 (172 eV).
Vertical line: position of H111.}
\end{figure}

In Fig. \ref{fig:spectrum2800} we compare harmonic spectra, on logarithmic scale, obtained
in different conditions. The black line indicates the spectrum produced in a very thin (20 $\mu$m) target with the IR pulse alone, which gives us practically
the single-atom spectrum, with the expected cutoff located at
H111. The conversion efficiency is very low due to the short cell.
The red line indicates the spectrum produced by the IR pulse recorded at the position in cell where we obtained the highest H111 intensity, i.e. -40 $\mu$m (cf. Fig. \ref{fig:chirped}(a)).
We observe that harmonics in the lower plateau (below 110 eV) experience an increase ($\approx 10^{2}$ %-10^{3}$
times) with cell length, as more atoms contribute to the HHG process, and apparently phase matching is favorable. For the higher plateau and cutoff harmonics we observe no yield enhancement with increasing cell length, which is explained by the PMM discussed in detail for H111.
The blue line depicts the spectrum obtained with the chirped THz pulse
assisting the HHG process at the position in the cell where the H111 intensity reaches its maximum (+260 $\mu$m).
Since the THz parameters were chosen to optimize QPM for H111, the spectrum is enhanced in that range. At H111 we obtain $\approx$ 1060 times increase
in the maximum harmonic yield after $\approx$1200 $\mu$m propagation compared to the yield at 20 $\mu$m.
With perfect PM and constant emission rate the harmonic yield should have increased quadratically with propagation distance, resulting in a $\approx$3600 times enhancement. Our QPM method reaches $\approx$30\% of this value while the emission rate is decreasing.
The effective harmonic enhancement in the selected spectral region is $\approx$ 100 times
using constant 120$\mu$m THz wavelength, and $\approx$ 300 times with chirped THz assistance.

%The first studies of THz assisted HHG were carried out with orders of magnitude higher THz field strengths and the main effect observed was the cutoff extension \cite{imre_pra2011, kinai}. The purpose of the co-propagating THz field was to increase the electric field strength in every other half-cycle.
In the present arrangement the THz pulse is too weak to increase the single-atom cutoff. For that orders of magnitude higher THz field strength is required [19, 22]. Here we induce QPM conditions to restore the single-atom cutoff and in addition
to enhance the yield
by around 300 times compared to the THz-free case.
%ADDED

In the geometry sketched in Fig. \ref{fig:sketch} for the used THz field strength and spot size we need 70 mJ THz pulse energy which is more than two orders of magnitude larger than available now in one-cycle THz pulse \cite{fulop}. However, using cylindrical focusing one can reduce the vertical THz spot size to 0.5 mm which lowers the needed THz energy by a factor of $\approx$10. By further optimization of the beam waist and the focusing in the cell the needed THz energy might be reduced to $\approx$4 mJ. Creating the THz field with the standing wave produced by two counter-propagating beams the needed energy can be reduced to $\approx$2$\times$1 mJ.
%There is more possibility to reduce the needed THz energy by optimization of the position of the THz spot and reducing the THz field without seriously sacrificing its enhancement effect.

Further investigation is going on to demonstrate that the presented QPM configuration is flexible
and that by tuning THz pulse parameters one can select the spectral range in which harmonic radiation can be amplified.
% ADDED
Our results demonstrate that by tuning the THz field one can finely shift the central frequency of the amplified spectral range toward lower harmonics. Another key is the IR pulse energy which shifts the HH cutoff. For our case, a $\pm$25\% IR pulse energy variation combined with appropriate THz field allowed amplification in the 140 -- 220 eV spectral domain.
% We can cut from the part below if we are in space deficite.
This might be of central importance in seeding free electron lasers in this range of photon energies. Recently mid-IR laser pulses are being also used for HHG in order to achieve higher cutoff energy. As the harmonic yield scales with $\lambda^{-5.5}$ \cite{tate}, an efficient QPM is even more important for this domain.
Applying the THz-assisted QPM scheme could
significantly increase the harmonic yield at selected spectral ranges approaching the water-window.

The authors acknowledge the support of: T\'AMOP-4.2.1/B-09/1/KONV-2010-0005 - Creating the Center of Excellence at the University of Szeged supported by the EU and co-financed by the European Social Fund, CNCS-UEFISCDI project no. PN-II-RU-PD-2011-3-0236 (KK); FP7 contract ITN-2008-238362 (ATTOFEL) (EB); Bolyai Postdoctoral Fellowship, Hungarian research grant OTKA 81364 (KV). We thank NIRDIMT Data Center and NIIF Institute for computation time.

\end{document}